\documentclass[prl,aps,twocolumn]{revtex4}
\usepackage{graphicx,amssymb,amsmath,color,psfrag}
\usepackage{amsthm}
\usepackage{amsfonts}
\usepackage{algorithmic}
\usepackage{enumerate}
\usepackage{latexsym}

\begin{document}
\newcommand{\dt}{\Delta\tau}
\newcommand{\al}{\alpha}
\newcommand{\ep}{\varepsilon}
\newcommand{\ave}[1]{\langle #1\rangle}
\newcommand{\have}[1]{\langle #1\rangle_{\{s\}}}
\newcommand{\bave}[1]{\big\langle #1\big\rangle}
\newcommand{\Bave}[1]{\Big\langle #1\Big\rangle}
\newcommand{\dave}[1]{\langle\langle #1\rangle\rangle}
\newcommand{\bigdave}[1]{\big\langle\big\langle #1\big\rangle\big\rangle}
\newcommand{\Bigdave}[1]{\Big\langle\Big\langle #1\Big\rangle\Big\rangle}
\newcommand{\braket}[2]{\langle #1|#2\rangle}
\newcommand{\up}{\uparrow}
\newcommand{\dn}{\downarrow}
\newcommand{\bb}{\mathsf{B}}
\newcommand{\ctr}{{\text{\Large${\mathcal T}r$}}}
\newcommand{\sctr}{{\mathcal{T}}\!r \,}
\newcommand{\btr}{\underset{\{s\}}{\text{\Large\rm Tr}}}
\newcommand{\lvec}[1]{\mathbf{#1}}
\newcommand{\gt}{\tilde{g}}
\newcommand{\ggt}{\tilde{G}}
\newcommand{\jpsj}{J.\ Phys.\ Soc.\ Japan\ }

\title{Controllability of ferromagnetism in graphene}
\author{Tianxing Ma$^{1,2,3,}$\footnote{txma@bnu.edu.cn}, Feiming Hu$^2$, Zhongbing Huang$^4$
and Hai-Qing Lin$^2$ } \affiliation{$^{1}$Department of Physics,
Beijing Normal University, Beijing 100875, China \\
 $^{2}$Department of
Physics and ITP,
The Chinese University of Hong Kong, Hong Kong \\
$^{3}$Max-Planck-Institut f\"ur Physik Komplexer Systeme,
N\"othnitzer Str. 38, 01187 Dresden, Germany\\
$^{4}$Faculty of Physics and Electronic Technology, Hubei
University, Wuhan 430062, China}

\begin{abstract}
We systematically study magnetic correlations in graphene within
Hubbard model on a honeycomb lattice by using quantum Monte Carlo
simulations. In the filling region below the Van Hove singularity,
the system shows a short-range ferromagnetic correlation, which is
slightly strengthened by the on-site Coulomb interaction and
markedly by the next-nearest-neighbor hopping integral. The
ferromagnetic properties depend on the electron filling strongly,
which may be manipulated by the electric gate. Due to its resultant
controllability of ferromagnetism, graphene-based samples may
facilitate the development of many applications.
\end{abstract}

\date{\today}
\maketitle

The search for high temperature ferromagnetic semiconductors, which
combine the properties of ferromagnetism (FM) and semiconductors and
allow for practical applications of spintronics, has evolved into a
broad field of materials science\cite{FE,FE1}. Scientists require a
material in which the generation, injection, and detection of
spin-polarized electrons is accomplished without strong magnetic
fields, with processes effective at or above room
temperature\cite{spintronics1}. Although some of these requirements
have been successfully demonstrated, most semiconductor-based
spintronics devices are still at the proposal stage since useful
ferromagnetic semiconductors have yet to be developed\cite{nano}.
Recently, scientists anticipate that graphene-based electronics may
supplement silicon-based technology, which is nearing its
limits\cite{Geim,Chen2009}. Unlike silicon, the single layer
graphene is a zero-gap two-dimensional (2D) semiconductor, and the
bilayer graphene provides the first semiconductor with a gap that
can be tuned externally\cite{Filling}. Graphene exhibits
gate-voltage controlled carrier
conduction\cite{Novoselov,Filling,VHS,bgate}, high field-effect
mobility, and a small spin-orbit coupling, making it a very
promising candidate for spintronics application
\cite{spintronics3,spintronics4}. In view of these characteristics,
the study of the controllability of FM in graphene-based samples is
of fundamental and technological importance, since it increases the
possibility of using graphene in spintronics and other applications.

On the other hand, the existence of FM in graphene is an unresolved
issue. Recent experimental and theoretical results in
graphene\cite{eet,HTc1,eebi} show that the electron-electron
interactions must be taken into account in order to obtain a fully
consistent picture of graphene. The honeycomb structure of graphene
exhibits Van Hove singularity (VHS) in the density of states (DOS),
which may result in strong ferromagnetic fluctuations, as
demonstrated by recent quantum Monte Carlo simulations of the
Hubbard model on the square and triangular lattices
\cite{hlubina,sqs}. After taking both electron-electron interaction
and lattice structure into consideration, the bidimensional Hubbard
model on the honeycomb lattice\cite{peres,Thereza,Rmp} is a good
candidate to study magnetic behaviors in graphene. Early studies of
the bidimensional Hubbard model on the honeycomb lattice were based
on mean field approximations and the perturbation theory\cite{Rmp}.
However, the results obtained are still actively debated because
they are very sensitive to the approximation used. Therefore, we use
the determinant quantum Monte Carlo (DQMC) simulation
technique\cite{dqmc,Hirsch} to investigate the nature of magnetic
correlation in the presence of moderate Coulomb interactions.
We are particularly interesting on
ferromagnetic fluctuations as functions of the electron filling,
because the application of local gate techniques enables us to
modulate electron filling\cite{Novoselov,Filling,VHS,bgate}, which
is the first step on the road towards graphene-based electronics.

The structure of graphene can be described in terms of two
interpenetrating triangular sublattices, A and B, and its low energy
magnetic properties can be well described by the
Hubbard model on a honeycomb lattice\cite{peres,Thereza,Rmp},
\begin{eqnarray}
H &=& -t\sum_{i\eta\sigma}a_{i\sigma}^{\dagger}
b_{i+\eta\sigma}+t'\sum_{i\gamma\sigma}
(a_{i\sigma}^{\dagger}a_{i+\gamma\sigma} +
b_{i\sigma}^{\dagger}b_{i+\gamma\sigma}) + {\rm h.c.}\nonumber \\
 &+& U\sum_{i}(n_{ai\uparrow}n_{ai\downarrow} +
n_{bi\uparrow}n_{bi\downarrow})+\mu\sum_{i\sigma}(n_{ai\sigma } +n_{bi\sigma })
\end{eqnarray}
where $t$ and $t'$ are the nearest and next-nearest-neighbor (NNN)
hopping integrals respectively, $\mu$ is the chemical potential, and
$U$ is the Hubbard interaction. Here, $a_{i\sigma}$
($a_{i\sigma}^{\dag}$) annihilates (creates) electrons at site ${\bf
R}_i$ with spin $\sigma$ ($\sigma$=$\uparrow,\downarrow$) on
sublattice A, $b_{i\sigma}$ ($b_{i\sigma}^{\dag}$) annihilates
(creates) electrons at the site ${\bf R}_i$ with spin $\sigma$
($\sigma$=$\uparrow,\downarrow$) on sublattice B, $n_{ai\sigma
}$=$a_{i\sigma}^{\dagger}a_{i\sigma}$ and $n_{bi\sigma
}$=$b_{i\sigma}^{\dagger}b_{i\sigma}$.

\begin{figure}
\includegraphics[scale=0.45]{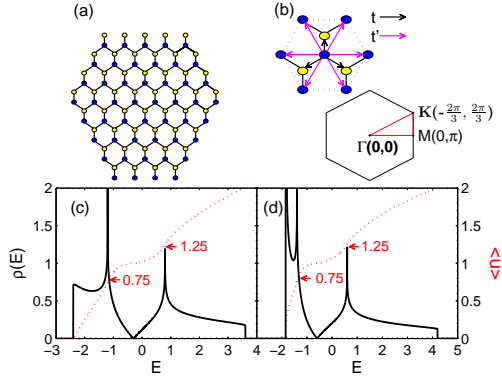}
\caption{(Color online) (a) Sketch of graphene with double-48 sites;
(b) First Brillouin zone and the high symmetry direction (red line);
(c) DOS (solid dark lines ) and fillings $\langle n\rangle$ (dash
red lines ) as functions of energy with $t'$=0.10t; and (d)
$t'$=0.20t. } \label{Fig:Structure}
\end{figure}

Our main numerical calculations were performed on a double-48 sites
lattice, as sketched in Fig. \ref{Fig:Structure}, where blue circles
and yellow circles indicates A and B sublattices, respectively. The
structure of the honeycomb lattice leads to the well known
massless-Dirac-fermion-like low energy excitations and the two VHS
in the DOS (marked in Fig. 1) at $<$$n$$>$=0.75 and 1.25
corresponding to $E$=$-2t'\pm t$, respectively as $t'<t/6$. While
when $t'\geq t/6$, a third VHS appears at the lower band edge, which
is a square root singularity marking the flattening of the energy
band near $\Gamma$ point.
They determine much of system's properties. According to the values
of $t$ and $U$ reported in the literatures for
graphene\cite{Rmp,U1,U2}, the ratio $U$/$\mid$t$\mid$ maybe expected
to be $2.2\sim 6.0$, which
is around the range of half-bandwidth to bandwidth\cite{bandwidth},
where the mean filed theory does not work well while the DQMC
simulation is a useful tool \cite{Hirsch}.
The exact value of $t'$ is not known but an {\it ab
initio} calculation \cite{Retal02} found that $t'/t$ ranges from
0.02 to 0.2 depending on the tight-binding parameterizations.
Therefore, it is necessary to study the ferromagnetic fluctuations
within the Hubbard model on the honeycomb lattice by including $t'$.

In the followings, we show that the behaviors of magnetic
correlation are qualitatively different in two filling regions
separated by the VHS at
$<$$n$$>$=0.75. In the filling
region below the VHS
the system shows a short-ranged
ferromagnetic correlation and the on-site Coulomb interaction tends
to strengthen ferromagnetic fluctuation. The ferromagnetic
properties depend on the electron filling, which may be manipulated
by the electric gate. Furthermore, the ferromagnetic fluctuation is
strengthened markedly as $t'$ increases. Our results highlight the
crucial importance of electron filling and the NNN hopping in
graphene. The resultant controllability of FM may facilitate the new
development of spintronics and quantum modulation.

To study ferromagnetic fluctuations, we define the spin susceptibility in the
$z$ direction at zero frequency,
\begin{eqnarray}
\chi(q) = \int_{0}^{\beta}d\tau \sum_{d,d'=a,b} \sum_{i,j}
e^{iq\cdot(i_{d}-j_{d'})} \langle\textrm{m}_{i_{d}}(\tau) \cdot
\textrm{m}_{j_{d'}}(0)\rangle
\end{eqnarray}
where $m_{i_{a}}(\tau)$=$e^{H\tau} m_{i_{a}}(0) e^{-H\tau}$ with
$m_{i_{a}}$=$a^{\dag}_{i\uparrow}a_{i\uparrow}-a^{\dag}_{i\downarrow}a_{i\downarrow}$
and
$m_{i_{b}}$=$b^{\dag}_{i\uparrow}b_{i\uparrow}-b^{\dag}_{i\downarrow}b_{i\downarrow}$.
Here $\chi$ is measured in unit of $\mid$$t$$\mid$$^{-1}$, and
$\chi(\Gamma)$ measures ferromagnetic correlation while $\chi(K)$
measures antiferromagnetic correlation.

\begin{figure}
\includegraphics[scale=0.375]{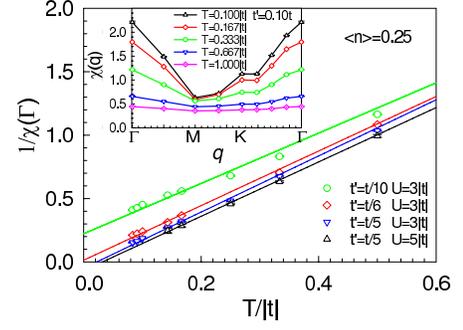}
\caption{(Color online) At $<$$n$$>$=0.25, inverse of magnetic
susceptibility, $1/\chi(q$=$\Gamma)$ versus temperature with
$U$=3$\mid$t$\mid$, $t'$=$t/10$, $t/6$, and $t/5$. Fitted line
$1/\chi(\Gamma)$=$\alpha$($T-\Theta$) are also shown. Inset:
Magnetic susceptibility $\chi(q)$ versus $q$ at different
temperatures with $t'$=0.10$t$ and $U$=3$\mid$$t$$\mid$.
}\label{Fig:Weiss}
\end{figure}

We first present temperature dependence of the magnetic correlations
at $<$$n$$>$=0.25 with different $t'$ and $U$. Fig. \ref{Fig:Weiss}
shows $1/\chi(q$=$\Gamma)$ versus temperature at
$U$=3$\mid$$t$$\mid$ with
$t'$=$t/10$, $t/6$, and $t/5$. Data for
$U$=5$\mid$$t$$\mid$ as $t'$=$t/5$ are also shown.
In the inset, we present $\chi(q)$ versus momentum $q$ at different
temperatures with $t'$=$t/10$ and $U$=3$\mid$t$\mid$. It is obvious
that $\chi(q)$ has strong temperature dependence and one observes
that $\chi(M)$ and $\chi(K)$ grow much slower than $\chi(\Gamma)$
with decreasing temperatures. Moreover, $1/\chi(\Gamma)$ exhibits
Curie-like behavior as temperature decreases from $\mid$$t$$\mid$ to
about 0.1$\mid$$t$$\mid$. Fitting the data as
$1/\chi(\Gamma)$=$\alpha$($T-\Theta$) (solid lines in
Fig.\ref{Fig:Weiss}) shows that
$\Theta$ is about 0.02$\mid$t$\mid$$\simeq 580K$ at $t'$=$t/5$, and
we also note that both $\Theta$ and $\chi(\Gamma)$ are enhanced
slightly as the on-site Coulomb interaction is increased. Positive
values of $\Theta$ indicate that the curves of $1/\chi(\Gamma)$
start to bend at some low temperatures and probably converge to zero
as $T\rightarrow 0$, $i.e.$, $\chi(\Gamma)$ diverges. This
demonstrates the existence of ferromagnetic state in graphene.

\begin{figure}
\includegraphics[scale=0.375]{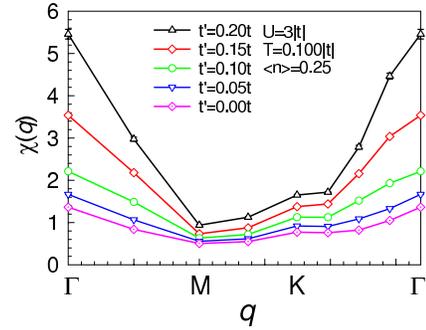}
\caption{(Color online) Magnetic susceptibility $\chi(q)$ versus
momentum $q$ at different $t'$, here $U$=3 $\mid$t$\mid$,
$<$$n$$>$=0.25 and $T$=0.10$\mid$t$\mid$. } \label{Fig:tp}
\end{figure}

From Fig. \ref{Fig:Weiss}, we may also notice that $t'$ plays a
remarkable effect on the behavior of $\chi(q)$, and results for
$\chi(q)$ dependent on $q$ with different $t'$ at
$U$=3$\mid$$t$$\mid$, $T$=0.10$\mid$$t$$\mid$ and $<$$n$$>$=0.25
have been shown in Fig. \ref{Fig:tp}. Clearly, $\chi(\Gamma)$ gets
enhanced greatly as $t'$ increases, while $\chi(M)$ and $\chi(K)$
increase only slightly. Thus, again it is significant to demonstrate
that ferromagnetic fluctuation gets enhanced markedly as $t'$
increases.
Furthermore, the strong dependence of FM on $t'$ suggests
controllability of FM in graphene by tuning
$t'$.

\begin{figure}
\includegraphics[scale=0.4]{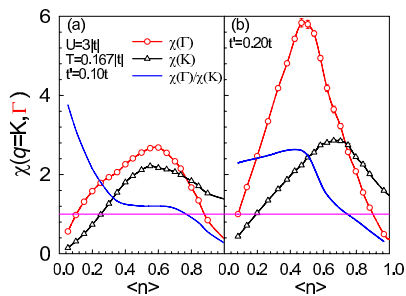}
\caption{(Color online) Magnetic susceptibility $\chi$($q$=$\Gamma$)
(red) and $\chi$($q$=$K$) (dark) versus electron filling at
$U$=3$\mid$t$\mid$ and $T$=0.167$\mid$t$\mid$ with (a) $t'$=0.1$t$
and (b) $t'$=0.2$t$.}\label{Fig:Filling}
\end{figure}


A great deal of current activity in graphene arises from its
technological significance as a
semiconductor material where
carrier density can be controlled by an external gate
voltage\cite{Novoselov,Filling,VHS}. To understand filling
dependence of magnetic correlations intuitively, we present
$\chi(\Gamma)$ (red), $\chi(K)$ (dark), and their ratio
$\chi(\Gamma)/\chi(K)$ (blue) versus filling for (a) $t'$=0.1$t$ and
(b) $t'$=0.2$t$ in Fig. \ref{Fig:Filling},
where $U$=3$\mid$$t$$\mid$ and $T$=0.167$\mid$$t$$\mid$. There is a
crossover between $\chi$($\Gamma$) and $\chi$($K$), which indicates
that the behaviors of $\chi$($q$) are qualitatively different in two
filling regions separated by the VHS at $<$$n$$>$=0.75, where the
ratio is 1. This is due to the competition between ferromagnetic and
antiferromagnetic fluctuations. The antiferromagnetic correlations
are strong around the hall-filing
and they may dominate the
shape of $\chi$($q$) in a wide filling range up to the VHS. The
effect of $t'$ in enhancing ferromagnetic fluctuation  can also be
seen by comparing Fig. \ref{Fig:Filling} (a) with (b). At electron
filling $<$$n$$>$=0.25, the ratio of $\chi(\Gamma)/\chi(K)$ is about
maximum for $t'$=0.2$t$ and is substantial for $t'$=0.1$t$, which is
the reason why did we choose electron filling 0.25 in Figs.
\ref{Fig:Weiss}, and
\ref{Fig:tp}. From the global picture
shown in Fig. \ref{Fig:Filling}, it is clear that the strength of
ferromagnetic correlation strongly depends on the electron filling,
which may be manipulated
by the electric gates in graphene, since $n\propto
V_{g}$\cite{Novoselov,Filling,VHS}. The filling region for
inducing FM required here likely exceeds the current experimental
ability. In fact, the challenge of increasing the carrier
concentration in graphene is indeed very important and it is a topic
now in progress. The second gate (from the top) and/or chemical
doping methods are devoted to achieving higher carrier
density\cite{bgate}.
Moreover, our results present here indicate the electron filling
markedly affects the magnetic properties of graphene, and the
controllability of FM may be realized in ferromagnetic
graphene-based samples. Furthermore, the change of ferromagnetic
correlation with
 $t'$ may also lead to controllability of FM in graphene. For example, one can tune $t'$ by
varying the spacing between lattice sites. Tuning $t'$ can also be
realized in two sub triangular optical
lattices\cite{Zhu2007,Wu2007}. Due to the peculiar structure of
graphene, which can be described in terms of two interpenetrating
triangular sublattices, making controlling $t^{\prime}$ in principle
possible in ultracold atoms system by using three beams of separate
laser\cite{Ruostekoski2009}.

In summary, we have presented exact numerical results on the
magnetic correlation in the Hubbard model on a honeycomb lattice. At
temperatures where the DQMC were performed, we found ferromagnetic
fluctuation dominates in the low electron filling region, and it is
slightly strengthened as interaction $U$ increases. The
ferromagnetic correlation showed strong dependence on the electron
filling and the NNN hoping integral. This provides a route to
manipulate FM in ferromagnetic graphene-based samples by the
electric gate or varying lattice parameters. The resultant
controllability of FM in ferromagnetic graphene-based samples may
facilitate the
development of many applications.


The authors are grateful to Shi-Jian Gu, Guo-Cai Liu, Wu-Ming Liu
and Shi-Quan Su for helpful discussions. This work is supported by
HKSAR RGC Project No. CUHK 401806 and CUHK 402310. Z.B.H was
supported by NSFC Grant No. 10674043.

\end{document}